\documentclass[vecphys]{svmult}

\usepackage{makeidx}         
\usepackage{graphicx}        
\usepackage{multicol}        
\usepackage[bottom]{footmisc}

\makeindex             

\begin{document}

\title*{The Extended Emission of Ultracompact HII Regions: An Overview and New Observations}
\titlerunning{The Extended Emission of UC~HII regions} 
\author{Eduardo de la Fuente Acosta\inst{1,2,3} \and
Stanley E. Kurtz\inst{2} \and
M. S. N. Kumar\inst{4} \and
Jos\'e Franco\inst{5} \and  
Alicia Porras\inst{1} \and
Simon N. Kemp\inst{3}}

\authorrunning{de la Fuente, Kurtz, Kumar, Franco, Porras \& Kemp} 
\institute{Instituto Nacional de Astrof\'\i sica \'Optica y Electr\'onica, M\'exico
\texttt{edfuente@inaoep.mx} \and
Centro de Radioastronom\'\i a y Astrof\'\i sica, UNAM, Morelia, M\'exico \and
Instituto de Astronom\'\i a y Meteorolog\'\i a, Dpto. de F\'\i sica, CUCEI, Universidad de 
Guadalajara, M\'exico \and
Centro de Astrofisica da Universidade do Porto, Portugal \and
Instituto de Astronom\'\i a, UNAM, M\'exico}
%
%
\maketitle

\begin{abstract}

\footnote{To appear in proceedings of the Puerto Vallarta Conference on ``New 
Quests in Stellar Astrophysics II: Ultraviolet Properties of Evolved Stellar 
Populations'' eds. M. Chavez, E. Bertone, D. Rosa-Gonzalez \& 
L. H. Rodriguez-Merino, Springer, ASSP series.

Presented as part of a Ph. D. thesis in the Departamento de F\'\i sica, of the Universidad de Guadalajara, M\'exico.}

  Ultracompact (UC) HII regions with Extended Emission (EE) are
  classical UC~HII regions associated with much larger ($>$1$'$)
  structures of ionized gas. The efforts to investigate, detect, and
  understand if the EE is physically related to the UC emission are
  few.  If they are related, our understanding of
  UC~HII regions may be affected (e.g., in the estimation of ionizing
  UV photons). Here we present a brief overview of UC~HII regions with
  EE (UC~HII+EE) including our most recent effort aimed at searching
  for UC~HII regions associated with extended emission.

\end{abstract}

\section{Introduction}
\label{sec:1}

Ultracompact (UC) HII regions are small (size $\leq$ 0.1 pc), dense
($\geq$ 10$^4$ cm$^{-3}$), photoionized hydrogen regions with high
emission measure ($\geq$ 10$^7$ {${\rm pc\ cm}^{-6}$}), surrounding
recently formed ionizing OB type stars (e.g. Fig.  1a). These
characteristics were observationally confirmed by \cite{WC89} and
\cite{K94}, and more recent reviews are presented by \cite{C02} and
\cite{LFR06}. The study of UC~HII regions began in 1967 via
interferometric observations of compact HII regions (see \cite{K02}
for a summary). Because UC~HII regions are generally surrounded by a
natal dust `cocoon', radio--continuum (RC) and infrared (IR)
observations are needed to study them. In the RC, the first VLA
surveys (e.g., \cite{WC89}, \cite{K94}) were made at 2 and 6 cm in
configurations A and B, supplying arc--sec resolutions and
sensitivities to structures up to 10--20$''$. The IR counterparts were
mainly provided by IRAS, with resolutions of $\sim$30$''$--2$'$.

\begin{figure}
\centering
\includegraphics[height=9.5cm]{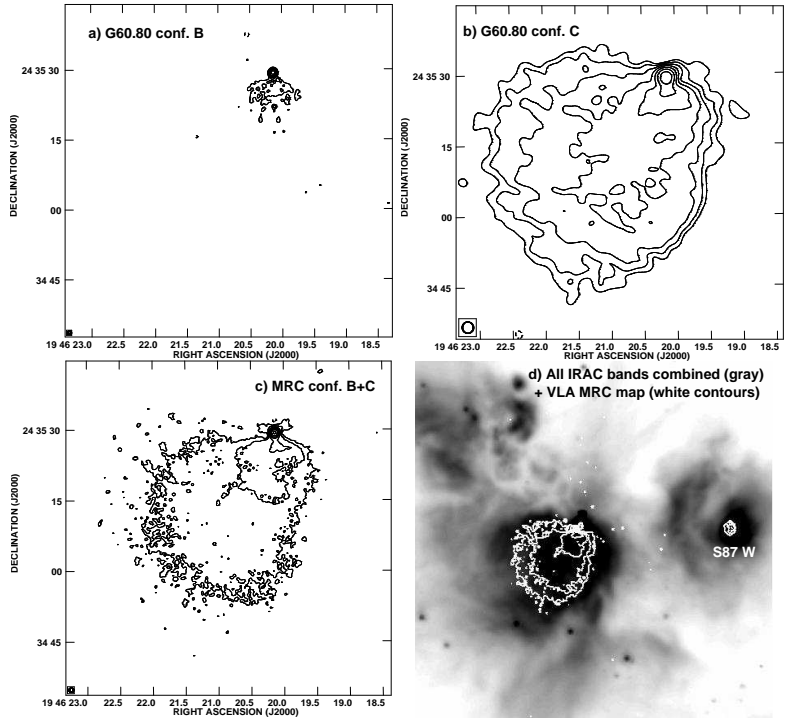}
%
%
\caption{The UC~HII region with EE G60.88--0.13 (IRAS 19442+2427). a) and b) VLA 
images in configuration B (the UC emission) and configuration C (the EE). c) Combined 
VLA Multi--Resolution--Clean (MRC) map. This map strongly suggests a direct connection 
between the UC and the extended emission. d) All IRAC bands image at meso--scales (gray) and 
superimposed contours from c). Dust is predominant in the region. A spatial location 
agreement between dust and the HII 
region is observed.}
\label{fig:1}       
\end{figure} 

Although the presence of large scale structures related to UC~HII
regions (Fig.  1b and 8c of \cite{K99}) had been inferred since 1967, the
first interferometric surveys did not detect them.  Nevertheless,
their detection is possible with the VLA C and D configurations (e.g.,
\cite{K99}), albeit at the expense of resolution towards the UC
emission (UCE).  To mitigate these spatial filtering effects, we made
multi--configuration VLA observations to provide a multi-scale view of
the UC~HII+EE (Fig.1c). Also, by using MSX observations, of higher
resolution than IRAS, it is possible both to detect the EE and
to resolve the UCE, since the infrared observations are sensitive
to the full range of
angular sizes. The best IR satellite observations available for this
purpose (see Fig. 1d) are from the Spitzer Space Telescope ({\it
  http://ssc.spitzer.caltech.edu}).

\section{Energetics and Extended Emission}

Comparing the total ionizing flux from the exciting star(s)
calculated using RC and FIR observations, it is possible to analyze
the energetics of the UC~HII regions. The spectral type of the exciting
star can be estimated in two ways: via the RC and via the IRAS fluxes.
In the former case, the ionizing photon rate, {\it N$'_c$}, is estimated
from VLA observations (eq. 1 of \cite{K94}) of the ionized gas.
In the latter case, the total luminosity is measured via IRAS fluxes,
and this luminosity is converted to an ionizing photon rate, N*$_c$,
using model stellar atmospheres (e.g., \cite{C86};
\cite{P76}).
Both estimates  consider an ionization--bounded, dust--free nebula.
N$'_c$ represents a lower limit to the spectral type because it is the
minimum flux required to maintain the observed UC~HII region; N*$_c$
represents an upper limit to the spectral type (for a single star producing
L$_{tot}$).

A more realistic case is the presence of a stellar cluster and also of
dust, as was considered by \cite{WC89} and \cite{K94}. Assuming a
spectral type for the most massive member of the cluster, and
comparing it with the spectral type based on N$'_c$ and  N*$_c$
(Table 18 in \cite{WC89} and Table 7
in \cite{K94}), they found that  N*$_c$ (IRAS fluxes) and a single-star 
assumption resulted in the earliest estimate for the spectral type.
The spectral type estimated from N$'_c$ was always of somewhat later
type.  The spectral type derived for the most massive member of a
cluster was roughly similar to that derived from the radio continuum.
This suggests that there is a significant amount of
dust in the regions, and many of the UV stellar photons do not
contribute to the ionization but heat the dust instead.
 
Another important result is that the luminosity
derived from IRAS suggests a much greater (earlier or more numerous)
stellar presence than does the VLA luminosity. This {\it IR--excess}
is quantified as f$_{d}$ = 1$-\xi$ = 1$-$(N$'_c$ / N*$_c$), where f$_d$
and $\xi$ are the fraction of UV photons absorbed by dust and gas
respectively. Values of f$_d$ $\sim$ 1 indicate a large IR--excess.
Surveys by \cite{WC89} and \cite{K94} found 0.42~$<$~f$_d$~$<$~0.99
for 29 sources.

On the other hand, the presence of EE in a {\it direct connection}
with the UC emission (i.e., a common structure of ionized gas embracing
both emissions; see Fig.  1c and Fig.  3 of \cite{K02}) would
significantly impact on our understanding of UC~HII regions. It may imply
that the definition, modeling, lifetime problem, and energetics of
UC~HII regions should be reconsidered. For example, as was pointed
out in \cite{K99} and \cite{K02}, if a direct connection is present,
and if we consider a single ionizing star, the EE requires
$\sim$~10--20 times more Lyman photons to maintain its ionization than
the needed by the UCE. So, if the Lyman photons are underestimated in the
N$'_c$ determinations of \cite{WC89} and \cite{K94}, then the role of
dust and the presence of clusters may have been over--estimated.

In light of the above, studies to understand and clarify the nature of
the EE are needed. To date, only four efforts have been made.

\subsection{The first three efforts}
 
The first effort was made by Kurtz et al. (1999; \cite{K99}) with a
sample of 15 UC~HII regions.  For 12 of them, they reported new,
low--resolution VLA
conf. D data at 3.6 cm (resolution of 9$''$ and sensitivity to 9$'$
structures). They contrasted these observations with those reported by
\cite{K94} (conf. B at 3.6 cm, resolution $\sim$ 0.9$''$; sensitivity
to about 10--20$''$) and the NVSS (conf. D at 20 cm, resolution
$\sim$~45$''$; sensitivity to about 7--15$'$). The aim was to detect a
direct connection between the UCE and the EE, via a morphological
study. If the UCE fits into a larger continuous structure, a direct
connection could be possible. If a falloff in the emission to a
near--zero value is observed between the UCE and the EE (a weak or null
contour in the map), the direct connection could be unlikely.  From
the 15 fields, they found EE in 12, and evidence for a direct
connection was present in eight. They calculate N$'_c$ using the
low--resolution VLA data and find that N$'_c$ is similar to N*$_c$.
 
The second effort was made by Kim \& Koo (2001; \cite{KK01}). This is
based on a previous single--object (G5.48--0.24) study (\cite{KK96}).
They studied 16 sources with the VLA conf. DnC at 20 cm with
resolution $\sim$ 30$''$ and sensitivity up to 15$'$. These
observations included RC and H76$\alpha$, He76$\alpha$
radio--recombination lines (RRL's). They found EE in all sources,
H76$\alpha$ in 15 and He76$\alpha$ in six.  Based on the RRL data,
they found similar LSR velocities in the UCE and EE, suggesting
a direct connection between the two. They also present a
theoretical model to explain the observed EE (density gradient
in the molecular cloud plus a champagne flow).
 
The third effort was made by Ellingsen et al. (2005; \cite{E05}).
They observed eight young (based on methanol maser emission) UC~HII
regions with the ATCA at 3.5 cm in the 750D configuration (resolution
and sensitivity similar to VLA conf. D at 3.6 cm). In a similar way as
\cite{K99}, they contrasted their observations with other ATCA high
resolution observations reported in the literature. They found EE with
smaller sizes than the ones in the sample of \cite{K99}, and explain that
this effect may be due to a younger sample selected. Also, they support
the theoretical model of \cite{KK01}.

\subsection{A new effort using VLA, 2MASS, and IRAC (Spitzer)}

The most recent effort was performed by de la Fuente (\cite{F07}) for
a sample of 29 UC~HII regions. The results will be published in a
series of forthcoming papers.  The aims were: (1) To complement and
confirm the results and assumptions of \cite{K99} with a sample of 14
regions (12 from \cite{K99} plus G35.20--1.74 and G19.60--0.23); and
(2) To explain how the presence of EE can affect the IR--excess
(f$_d$), in the complete sample that includes 15 regions in
\cite{WC89} and \cite{K94}.  Values of f$_d$ $>$ 0.8 were
found for this UCE sample.

The methodology employed includes, when available, NIR (2MASS), MIR
(IRAC \cite{F04} from the GLIMPSE program, {\it
  http://www.astro.wisc.edu/sirtf/}), and RC (VLA at 3.6 cm in
configurations D, C, and B) observations. RC traces the
ionized gas while NIR and MIR are good tracers of the stellar population
and dust, respectively.  With IRAC it is possible to detect dust and
PAH emission at 3.6, 5.8 and 8.0~$\mu$m. Shocked gas can be observed 
through H$_2$ emission at 4.5~$\mu$m. The 3.6~$\mu$m (Spitzer)
and the K$_s$ (2.12~$\mu$m, 2MASS) images can show the presence of
stellar clusters and nebulosities (ionized or reflection nebula). With
IR photometry of these data, it is possible to identify the
YSO population. Here, we summarize only the results regarding IRAC
imagery and VLA maps.

Combining new VLA observations in conf. C with previous data in conf.
B and D, a Multi--Resolution--Clean (MRC) map was created (e.g., Fig.
1c). If this map shows that the extended and UC emission form part of
a continuous structure, then the suggestion of a direct connection is
strong.

This behavior was confirmed in seven of the 14 regions in our sample.
It was not seen in two (G33.13--0.09 and G48.61+0.02), and it is
probable in five, although other observations (e.g., IRAC) are needed to
confirm. In general, the results of \cite{K99} were confirmed for some
of the sources, however, further observation and analysis is required
for the others. For
example, for G78.44+2.66 and G106.80+5.31, the MRC maps do not confirm
a direct connection between the UCE and the EE (in agreement with
\cite{K99}), nevertheless, the K$_s$ image shows a nebulosity covering
both emission regions. IRAC images (not currently available) could confirm
the nature of the nebulosities.


Using  new VLA conf. D observations at 3.6 cm for the 15 sources
to complete the sample of 29, a determination of f$_d$ was computed.
Also in these new RC maps, EE is present in all sources.  Based on
their morphology, and following \cite{K99}, a direct connection is
also strongly suggested in 12 of the 15 regions. The cometary
morphology was predominant in the whole sample.
Confirming that the 29 sources
present EE and IR--excess, a comparison between f$_d$ (conf. D) and
f$_d$ (conf. B) was performed. For all sources, the presence of EE
reduces the values of f$_d$. In summary, 10 regions have f$_d$ (conf.
D) $\sim$~0.2, another 10 regions $\sim$~0.6, and the other 9,
$\sim$~0.7.

The PAH emission is a good tracer of the ``radiation temperature'' and
the IRAC 8 $\mu$m band is dominated by this emission (predominant at
7.7~$\mu$m). The striking comparison between the IRAC images and the
VLA RC images (see Fig.1d) suggests that the EE is due to ionizing
radiation.  However, soft UV radiation which may not significantly
contribute to the overall HII region could be an important ionization
source for the EE.  In the 8.0~$\mu$m image, several knot--like
sources are observed. They could be either star clusters or
externally illuminated condensations.  Furthermore, the 8.0~$\mu$m
band can effectively trace weak structures and has proven useful in
unveiling the underlying physical structure of the dense core/cloud
(e. g., \cite{He07}, \cite{KG07}).
Also, an agreement between the location of the RC emission peaks
in the cometary arcs and the strongest emission in IRAC bands was
observed.

In summary, the EE seems to be common in UC HII regions and is
deserving of special attention in forthcoming studies and analysis.
Multi-configuration VLA maps are critical to study morphologically the
direct connection between UC~HII regions and their associated EE.
IRAC has been revealed as a powerful tool to study UC~HII regions
(with or without EE). The EE helps to explain the IR--excess
observed because the N$'_c$ calculated in \cite{WC89} and \cite{K94}
under--estimated the ionizing Lyman photons. Nevertheless, the
over--estimation of dust in the regions is not necessarily true. On the
other hand, in several sources the presence of clusters of stars in
the UC~HII+EE is inferred, in agreement with \cite{K94}. Hence,
the assumptions of a single ionizing star and dust free nebula are not
necessarily valid in the energetic studies.

\section{The Aftermath}

All the efforts are complementary, and individually none of them
clarify the whole scenario. More studies are needed to clarify the nature,
formation and evolution of the EE. A starting point is to
standardize the observations (Molecular, HI, VLA multi--configuration,
RRL's, and Spitzer) of all sources presented in these efforts. Combining
these observations it is possible: 1) To confirm the validity of the
model presented by \cite{KK01}.  2) To measure the extinction. If similar extinction is measured in the UCE and EE, then it is possible to
guarantee a direct relation between these components. This could be
done by comparing NIR recombination lines with RC images.
  3) To compare HI, RC, and IRAC
observations looking for a relation between HII regions and PDR's. \\

E de la F very gratefully thanks CONACyT (grants 124449 and SNI III 1326 M\'exico) for financial support from CONACyT . He also acknowledges support from Centro de Astrofisica da Universidade do Porto and Universidad de Guadalajara (CUCEI) during his stay in Portugal.


%
%



%
%
%
%
%

%
%
%
%
%
%
%
%


\printindex
\end{document}